\documentclass[
nosuperscriptaddress, twocolumn,
amsmath,amssymb,aps, nofootinbib,
floatfix, prl ]{revtex4-1}
\usepackage{graphicx}
\usepackage{dcolumn}
\usepackage{bm}
\usepackage{amsmath,amssymb,mathrsfs,amsfonts}
\usepackage{hyperref,xfrac,color,xcolor}

\newcommand{\p}{\bm{p}}
\newcommand{\q}{\bm{q}}

\newcommand\supplemental

\begin{document}

\title{Azimuthal Anisotropy at high transverse momentum in $p$-$p$ and $p$-$A$ collisions }

\author{Ismail Soudi}
\email{ismail.soudi@wayne.edu}
\affiliation{
	Department of Physics and Astronomy, Wayne State University, Detroit, MI 48201. }
\author{Abhijit Majumder}%
\email{majumder@wayne.edu}
\affiliation{
	Department of Physics and Astronomy, Wayne State University, Detroit, MI 48201. }

\date{\today}

\begin{abstract}
	We explore the possibility that the initial transverse momentum distribution of unpolarized and polarized partons within unpolarized nucleons, both with and without the anisotropy of unpolarized hadrons produced in the fragmentation of outgoing partons, could lead to the observed azimuthal anisotropy of high transverse momentum (high-$p_T$) hadrons produced in high energy proton-proton ($p$-$p$) or proton-ion ($p$-$A$) collisions. Including simple Gaussian forms for transverse momentum dependent parton distribution functions (PDF) and fragmentation functions, and assuming an $A^{1/3}$ enhancement of a PDF in $p$-$Pb$ collisions, we show that the observed anisotropy, with \emph{no modification} to the angle integrated spectra ($R_{pA}\!=\!1$) for 5~GeV~$\!\lesssim\! p_T\!\lesssim\! 50$~GeV, can be straightforwardly understood as arising from a few processes dominated by gluon-gluon to gluon-gluon scattering. 
\end{abstract}

\maketitle

\emph{Introduction:} Azimuthal anisotropy of hadrons with transverse momentum $p_T \lesssim 3$~GeV~\cite{STAR:2004jwm,PHENIX:2003qra,ALICE:2010suc,CMS:2012zex,ATLAS:2012at}, in a semi-central relativistic heavy-ion collision, consistent with calculations of an expanding elliptically asymmetric fluid dynamical system, with a small specific shear viscosity [$\eta \lessapprox s/(2\pi)$, where $s$ is the entropy density] is  routinely proffered as evidence for the formation of a quark gluon plasma~(QGP)~\cite{Romatschke:2007mq,Shen:2011eg,Song:2010mg,JETSCAPE:2020mzn,JETSCAPE:2020shq}.
The high densities (and associated gradients) that cause the collective flow of the QGP, inescapably require the modification (energy-loss) of high-$p_T$ partons (quarks, gluons) traversing the plasma~\cite{Bjorken:1982tu,Gyulassy:1993hr,Baier:1996sk,Zakharov:1997uu,Gyulassy:2000er, Wang:2001ifa,Arnold:2002ja}. This \emph{jet quenching} is discerned via the modification of high-$p_T$ jets~\cite{ATLAS:2012tjt,CMS:2011iwn,ALICE:2013dpt} and hadrons~\cite{ATLAS:2015qmb,CMS:2012aa,ALICE:2010yje,STAR:2003fka,PHENIX:2001hpc} in central and semi-central heavy-ion collisions. Successful comparisons with simulations, which utilize the expanding medium profile, required for the bulk soft spectrum~\cite{JETSCAPE:2022jer}, are considered a corroborative signature of the QGP.

It is straightforward to expect that for semi-central collisions, where the bulk medium may possess a component of elliptic deformation in the directions transverse to the colliding beams, the quenching of jets in these directions will also be different~\cite{Wang:2003mm,Majumder:2006we,Bass:2008rv}. A large,  centrality dependent, suppression of jets or high-$p_T$ hadrons should go hand-in-hand with an azimuthal anisotropy in the modification of these jets or hadrons~\cite{Noronha-Hostler:2016eow,Kumar:2019uvu,He:2022evt}. This is indeed what is observed in semi-central heavy-ion collisions.

Recently, several measurements of high multiplicity $p$-$p$ and $p$-$Pb$ events~\cite{ATLAS:2017hap,ALICE:2014dwt,CMS:2015yux} have demonstrated the existence of an azimuthal anisotropy in the soft sector ($p_T \!\lesssim\! 3$~GeV), revealed in terms of the Fourier coefficients of the azimuthal distribution of the yield of particles:
\begin{eqnarray}
	\frac{dN}{d\phi} &\propto&  1 + \sum\limits_{n=1}^{\infty} 2 v_n \cos\left[ n( \phi - \Psi_n ) \right].
\end{eqnarray}
In the equation above, $v_n$ are the $n^{\rm th}$ harmonic flow coefficients with respect to the flow plane angle $\Psi_n$. While the existence of a $v_n$ in these \emph{small systems}, comparable to those in $Pb$-$Pb$ collisions, are indicative of the formation of droplets of QGP in these systems, no (angle integrated) modification of high-$p_T$ jets~\cite{ALICE:2015umm,CMS:2016svx} or hadrons~\cite{ATLAS:2022iyq,ALICE:2021est} has been found in these systems. Yet, $p$-$Pb$ (and even $p$-$p$) events show an azimuthal anisotropy for hadrons at high-$p_T$ (measured up to $50$~GeV in $p$-$Pb$), 
~\cite{ATLAS:2014qaj,ATLAS:2016yzd,ATLAS:2019vcm}. This indicates that a mechanism other than energy loss could be the source of this asymmetry.

In this Letter, 
azimuthal anisotropies in high-$p_T$ hadron production will be shown to arise from the asymmetric hard scattering between incoming partons which possess a small transverse momentum relative to the $z$-axis, defined by the opposing proton beams~\cite{Field:1977ha}. These Transverse Momentum-dependent Distributions (TMDs)~\cite{Collins:1981uw,Collins:1981va,Collins:1981zc} for both the incoming and outgoing hadrons will display further angular asymmetries in the case when the quarks possess transverse spin~\cite{Boer:1997nt} or the gluons possess linear polarizations~\cite{Mulders:2000sh}. The largest contributions emanate from $g+g\rightarrow g+g$ scattering with polarization independent gluon ($g$) TMDs, and from combinations of parton distribution functions (PDFs) with a linearly polarized gluon within un-polarized protons (the Boer-Mulders' effect~\cite{Boer:1997nt,Mulders:2000sh}), followed by polarization dependent hard scattering and transverse momentum dependent fragmentation to an un-polarized hadron from a linearly polarized gluon (the Collins' effect~\cite{Collins:1992kk}).

The presence of $\cos{ 2 \phi }$  asymmetries in the production of hadrons involving the linear polarization of gluons (or transverse quark spin) is well known in a variety of processes~\cite{Collins:1992kk,Gamberg:2005ip,Conway:1989fs,NA10:1986fgk,EuropeanMuon:1983tsy, Barone:2009hw, Boer:2010zf, Pisano:2013cya}. In this Letter, we will demonstrate that these $\cos{ 2 \phi }$  asymmetries, extended to $p$-$p$~\cite{Pisano:2013cya,Boer:2010zf,Anselmino:2004ky,Anselmino:2005sh}, lead to the elliptic anisotropy ($v_2$) observed in $p$-$p$ and $p$-$Pb$, without modification of the angle integrated spectrum. 

This Letter is organized as follows: We review the formalism for 2-to-2 scattering of hard partons at leading order leading twist, where the initial and final hadrons are unpolarized. The intermediate partonic states may, or may not, include a transverse spin-polarized quark and/or a linearly polarized gluon. We then outline the phenomenological setup to calculate the differential angle dependent cross section and outline how a $v_n$ can be extracted from these calculations. This is followed by comparison of our calculations with data from $p$-$p$ and $p$-$A$ collisions. While our calculations will directly apply to the case of high-$p_T$ hadron production in $p$-$p$ collisions, we will make phenomenological extensions to the case of $p$-$A$ collisions (a rigorous calculation of spin-polarization dependent multiple scattering is yet to be carried out).

\paragraph{Theoretical framework:}
There have been extensive studies of high-$p_T$ hadron production in $p$-$p$ collisions, where the intrinsic transverse momentum of partons within the incoming proton~\cite{Boer:2010zf,Pisano:2013cya,denDunnen:2014kjo,Boer:2014tka}, and the transverse momentum of the final hadron with respect to the high-$p_T$ parton~\cite{Bacchetta:1999kz,Anselmino:2004ky,Anselmino:2005sh,DAlesio:2004eso} have been included. Following \cite{Anselmino:2004ky,Anselmino:2005sh}, the differential cross section for an un-polarized $p$-$p$ collision leading to the production of an un-polarized hadron [a pion ($\bm{\pi}$) with rapidity $y$ and transverse momentum $P_T$], but including internal partonic states that may be polarized, can be expressed as, 
\begin{eqnarray}
	&\frac{d\sigma}{dy d^2 \!P_{T}} \!\!
	=\!\!\int\!\! \frac{dx_a dx_b dz d^2 k_{\!\bot a} d^2 k_{\!\bot b} d^3 k_{\!\bot C} }{2 \pi^2 z^3 s} \delta(\bm{k}_{\!\bot C}\!\cdot\!\hat{p}_c)J(\bm{k}_{\!\bot C}) \label{eq:cross-section} \:\:\:\;\;\\
	&  \Gamma^{\sigma\mu}\!(x_a,\!k_{\!\bot a}\!) \Gamma^{\alpha \nu}\!(x_b,\!k_{\!\bot b}\!) 
	\hat{M}_{\mu\nu\rho} \hat{M}^{*}_{\sigma\alpha\beta}  \Delta^{\rho\beta}\!(z,k_{\!\bot C}\!) \delta(\hat{s} + \hat{t} + \hat{u}) , \nonumber
\end{eqnarray}
where $J(\bm{k}_{\bot C}) = \frac{(E_C^2+\sqrt{\bm{p}_C^2-\bm{k}_{\bot C}^2})^2}{4(\bm{p}_C^2-\bm{k}_{\bot C}^2)}$.
All partons are massless (ensured by the final $\delta$-function). The fourth parton is un-hadronized allowing for any final state. A sum over all flavors is implied, and the integral is over unconstrained components of the momenta.
While the generalized PDF $\Gamma^{\mu\nu}$ and fragmentation function (FF) $\Delta^{\mu\nu}$ may describe quarks or gluons, in this Letter, we will focus on the gluon-gluon partonic channel which dominates the cross section for pion production at the $p_T$'s and $\sqrt{s}$ considered.
The generalized gluon PDF reads \cite{Mulders:2000sh}
\begin{eqnarray}
	{\mkern-36mu}
	\Gamma^{\mu\nu}_{P}(x,k_{\bot})
	%
	%
	&= \frac{
	-g_T^{\mu\nu} f(x,k_{\bot}) 
	+ \left(\frac{k_{\bot}^{\mu}k_{\bot}^{\nu}}{M_p^2} + g_T^{\mu\nu}\frac{k_{\bot}^2}{2M_p^2} \right)h^{\bot}(x,k_{\bot}^2)}{2x}\;,
	&{\mkern-36mu}
\end{eqnarray}
with $g_{T}^{\mu\nu}\!=\! g^{\mu\nu} \!\!- n^{\mu}_+ n^{\nu}_- \!\! - n^\mu_- n^\nu_+$.
The $f(x,k_\perp^2) $ represents spin-polarization independent TMD-PDFs for each incoming parton ($a,b$), with longitudinal momentum fraction $x_{(a,b)}$ and transverse momentum $k_{\bot (a,b)}$, relative to the $z$-axis defined by the incoming beams.
The distribution of linearly polarized gluons in nucleons is given by the Boer-Mulders' function $h_1^{\bot}(x,k_{\bot}^2)$ \cite{Boer:1997nt}.
Using gluon polarization vectors, at leading order in transverse momentum, given by
$\epsilon^{\lambda}_{\mu} = \sfrac{e^{-i\lambda\phi} \left(0,-\lambda, -i,0\right)}{\sqrt{2}} + \mathcal{O}\left(\sfrac{k_{\bot}}{p_z}\right)$, one can project the gluon into correlated helicities,
\begin{eqnarray}\label{eq:CorrelatorPDF}
	&\mbox{}\!\!\!\!\Gamma^{\lambda_1,\lambda_2}_P
    (x,k_{\bot}\!) \!
	= \!\frac{
	-\delta^{\lambda_1,\lambda_2} f
    (x,k_{\bot}^2\!) 
	+\delta^{\lambda_1,-\lambda_2}\frac{k_{\bot}^2}{2M_p^2} h^{\bot}
    (x,k_{\bot}^2\!)
    }{2x}
    .&
\end{eqnarray}

Similarly, for the fragmentation, the correlator is, 
\begin{eqnarray}\label{eq:CorrelatorFF}
	&\!\!\!\Delta^{\lambda_1,\lambda_2}
    (z,k_{\bot}\!) 
	\! =\! \frac{
	\!-\delta^{\lambda_1,\lambda_2} D(z,k_{\bot}^2\!) 
	+\delta^{\lambda_1,-\lambda_2}\frac{k_{\bot}^2}{2M_{\pi}^2} H^{\bot}(z,k_{\bot}^2\!)}{2/z}.&
\end{eqnarray}
Here $D(z_c,k_{\bot C})$ represents the spin-polarization independent TMD-FF for the outgoing parton (c) fragmenting to the detected $\bm{\pi}$, carrying momentum $z \bm{p}_c + \bm{k}_{\bot C}$ (the first $\delta$-function in Eq.~\eqref{eq:cross-section} ensures that $\bm{k}_{\bot C}$ is perpendicular to the 3-momentum of the outgoing parton $\hat{p}_c$).
The distribution of fragmenting $\bm{\pi}$ from a linearly polarized gluon is given by the Collins' function $H^{\bot}(z,k_{\bot})$ \cite{Collins:1992kk}.

Using the helicity projected correlators, the combination of matrix element times complex conjugate with initial and final correlators can be written as
\begin{equation}
	\Sigma \equiv
	\sum\mbox{}_{\mbox{}_{\{\lambda_i\}}}
	\Gamma^{\lambda_a,\lambda_a'} \Gamma^{\lambda_b,\lambda_b'}
	\hat{M}_{\lambda_a , \lambda_b}^{\lambda_c , \lambda_d} \left(\hat{M}_{\lambda_a', \lambda_b'}^{\lambda_c', \lambda_d'}\right)^*\!\!  \Delta^{\lambda_c,\lambda_c'}\;. \label{eq:sigma}  
\end{equation}
The $\hat{M}_{\lambda_a , \lambda_b}^{\lambda_c , \lambda_d}$ represents the matrix element for helicity dependent hard scattering, where the helicities could be different in the complex conjugate.
To obtain the familiar formula for spin-polarization independent scattering, one sets each $\lambda' = \lambda$, and sums over the helicities, to obtain, 
\begin{eqnarray}
& \!\!\!\!\! \frac{d\sigma^0}{dy d^2 P_{T}}\!\!
	=\!\!\int\!\! \frac{dx_a dx_b dz d^2k_{\bot a} d^2 k_{\bot b} d^3 k_{\bot C} }{16 \pi^2 x_a x_b z^2 s}  \delta(\bm{k}_{\bot C}\!\cdot \!\hat{p}_c)J(\bm{k}_{\bot C}) \label{eq:cross-section-spin-independent}\\
 & \!\!\!\!\! \hat{f}_{\!\mbox{}_{a/A}}\!(\!x_a,\! k_{\bot a}\!) 
	 \hat{f}_{\!\mbox{}_{b/B}}\!(\!x_b,\! k_{\bot b}\!) 
  | M_{a,b}^{c,d} |^2  \hat{D}^{\mbox{}^{C/c}}\!(\!z,k_{\bot C}\!) \delta(\hat{s} + \hat{t} + \hat{u}). \nonumber
\end{eqnarray}
While the equation above was derived for gluon-gluon scattering, it also  holds for all other partonic scatterings, with a simple replacement of different distributions.

Eqns.~(\ref{eq:cross-section},\ref{eq:cross-section-spin-independent}) may be combined: $\sfrac{d \sigma}{[dy d^2 P_T]} \!\!=\! \sfrac{d \sigma^0}{[dy d^2 P_T]} \, + \, \sfrac{d \Delta \sigma}{[dy d^2 P_T]}$. The $\Delta \sigma$ explicitly refers to contributions where the helicity of any parton in the amplitude is opposite to its helicity in the complex conjugate. 
Both $d \sigma^0$ and $d  \Delta \sigma$ will contribute to azimuthal anisotropies.

\begin{figure}[!htb]
	\begin{center}
		\includegraphics[width=0.4\textwidth]{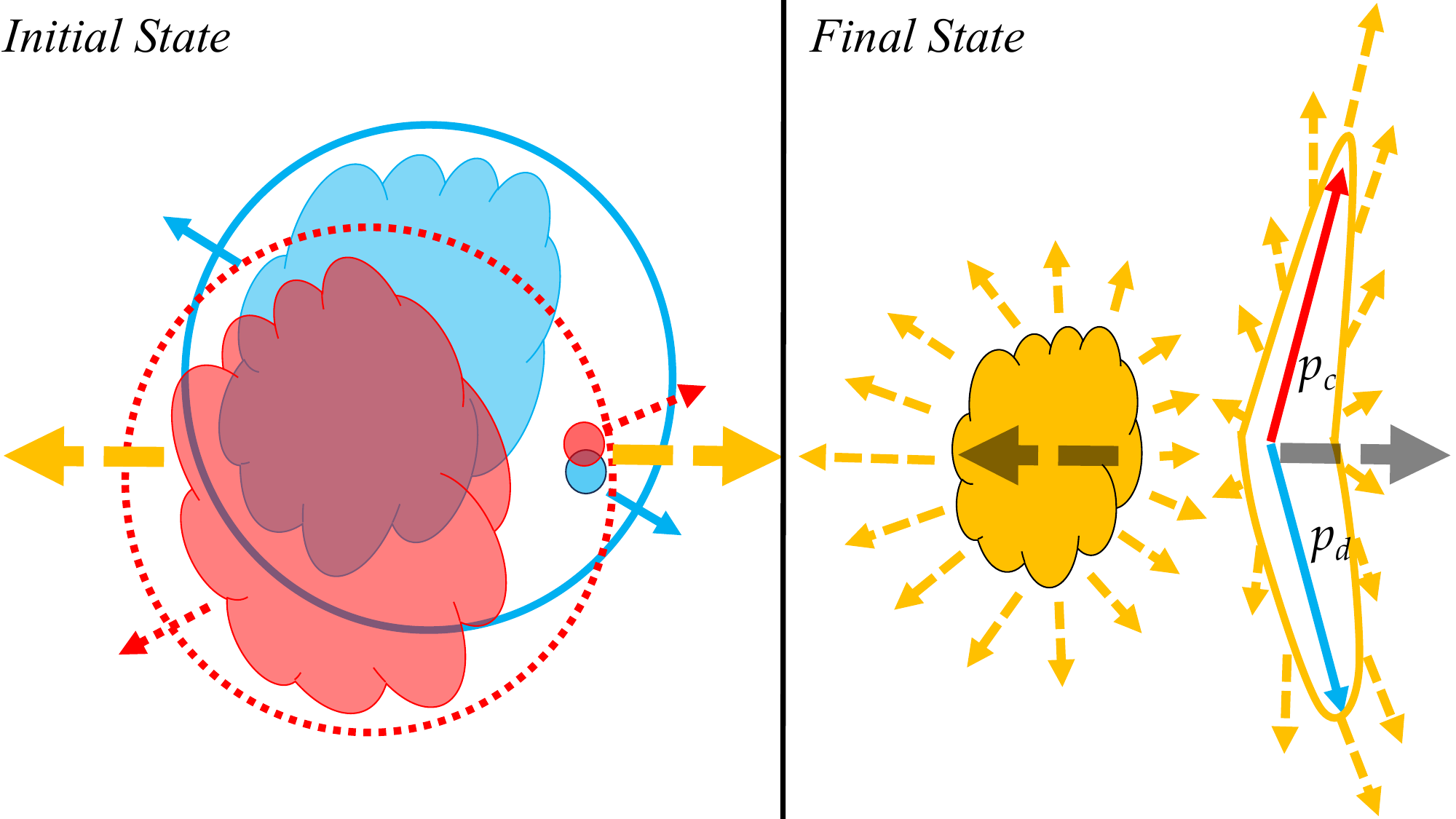}
	\end{center}
 \vspace{-0.6cm}
	\caption{Color Online: Left: Illustration of the initial state of a $p$-$p$ collision. Blue proton with a solid boundary is coming out of page, red proton with dotted boundary goes into page. Hard parton is shown as a small dot (blue/red) with a small transverse momentum indicated with small arrow, balanced by the transverse momentum of the artificially separated remaining matter, indicated by the blob. Right: Net transverse momentum of the final state hard di-jet balanced with the net momentum of the produced remainder (gray arrows).}
	\label{fig:v2-illustration}
\end{figure}

\paragraph{Azimuthal Anisotropy:}
Consider a hard parton which carries a non-negligible fraction $x_a$ of the momentum of proton $A$ and a transverse momentum $k_{\bot, a}$. This is indicated as the solid blue dot in Fig.~\ref{fig:v2-illustration}. The remainder of the nucleon (bounded by outer blue circle) is indicated by the blue blob and contains both hard partons and soft bulk. Due to momentum conservation, the net transverse momentum of the remainder is  $-k_{\bot, a}$ (opposing solid blue arrows in the figure). For a hard 2-to-2 scattering, a similar configuration must arise in the opposing proton indicated by the red dashed boundary in the figure: The transverse momentum of a hard parton $k_{\bot, b}$ is balanced by the remainder (red blob in Fig.~\ref{fig:v2-illustration}), and indicated by the dashed red arrows. The net momentum of the participating hard partons is balanced by the remainders, indicated by the opposing yellow arrows.

The hard partons scatter to produce 2 back-to-back jets in the right panel of Fig.~\ref{fig:v2-illustration}. 
The entire remainder of the protons does not interact in the collision. The portion that does, is indicated as the orange blob in the right panel of Fig.~\ref{fig:v2-illustration}. 
The net momentum of the remainder does not exactly balance that of the di-jets. However, there will be a strong correlation between the opposing net momentum of the interacting remainder and the participating di-jets (the di-jet system is deflected towards the right and the bulk particle production is boosted towards the left). Soft hadrons are produced in the hadronization of the remainder, as well as from the fragmentation of the jets. The correlated opposing momenta of the two systems will form an approximate \emph{event plane}. 

The above picture can be realized with a PYTHIA simulation~\cite{Bierlich:2022pfr}: A single hard scattering event without initial or final state parton shower is set up. 
The Lund model relies on beam remnants, which are the partons that do not participate in the hard scattering but continue traveling along the $\pm z-$direction.
For each remnant we sample a transverse momentum $\bm{k}_{\perp i}$, following a Gaussian distribution with an average momentum square $\langle k_\perp^2\rangle$ and random transverse direction.
The total transverse momentum of all remnants $\bm{q}_T=\sum_i \bm{k}_{\perp i}$ is compensated by applying a Lorentz boost to the center of mass frame of the hard scattering with the velocity 
$    \bm{v} = \sfrac{\bm{q}_T}{\sqrt{(E_a + E_b)^2 + \bm{q}_T^2}}$.
By applying a boost to the center of mass frame, we ensure that the scattering cross-section is not altered. 

\begin{figure}
    \centering
    \includegraphics[width=0.475\textwidth]{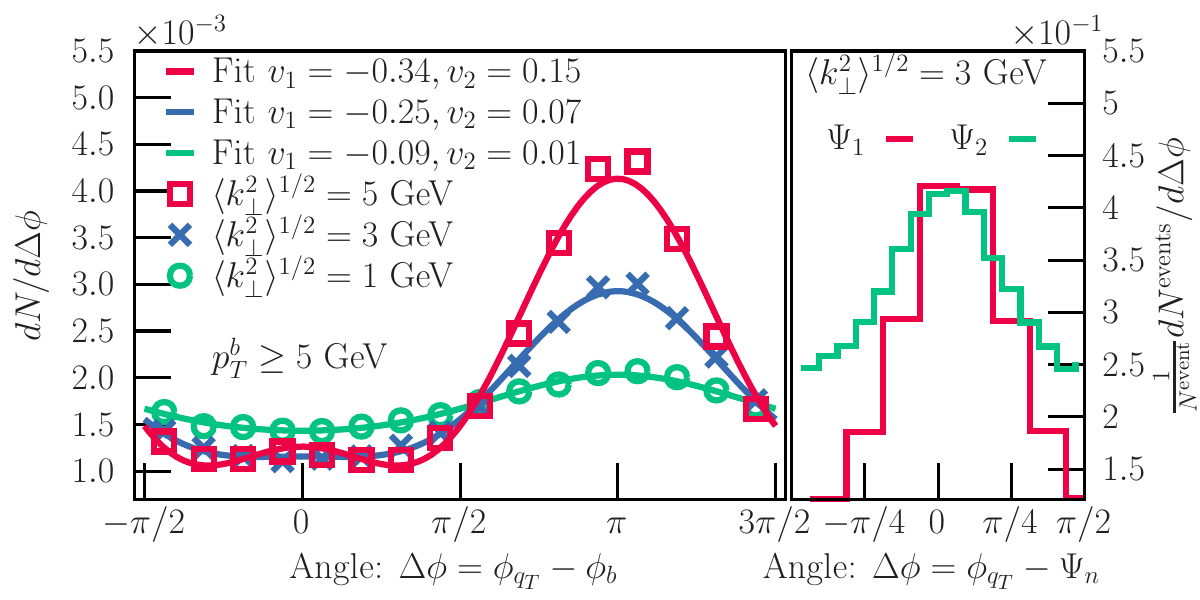}
    \vspace{-0.3cm}
    \caption{Left: Template fits [Eq.~\eqref{eq:template}] to the azimuthal distribution of hadrons with $p_T\! \geq\! 5$~GeV with respect to the event angle $\phi_{\q_T}$ set by the vector $\vec{q}_T$ (see text). Three different choices of the mean intrinsic transverse momentum $\langle k_\perp^2 \rangle^{1/2}$ are presented: $1$~GeV is the value used for $p$-$p$ collisions in Fig.~\ref{fig:v2Band}, progressively higher values of $3,5$~GeV visually illustrate the escalation of an elliptic anisotropy. Right: The event fraction distribution of the difference between the direction of the di-jet angle $\phi_{\q_T}$ and the event plane angles $\Psi_n$ (see text). }
    \label{fig:template_event_plane_correlation}
\end{figure}

After hadronization, using the $Q$-vector method~\cite{Borghini:2001vi} with the azimuthal angles $\phi_j$ of the outgoing hadrons, we determine the $n^{\rm th}$-order event plane $\Psi_n\!\!=\!\arg [\sum_{j} \! e^{i n \phi_j}\!]/n$.
The sum over $j$ includes all hadrons in the event. The $\Psi_n$ are found to be tightly correlated with the direction of the $\q_T$ vector, $\phi_{\q_T}$, as demonstrated in the right panel of Fig.~\ref{fig:template_event_plane_correlation}, which shows the distribution of event fraction for the difference between the $\phi_{\q_T}$ and the first or second order event plane $\Psi_1, \Psi_2$, for $\langle k_\perp^2 \rangle^{1/2}\! =\! 3$~GeV.

A high-$p_T$ azimuthal anisotropy will result if there is an angular ($\phi$) correlation between the direction of the highest $p_T$ hadrons, projected on the azimuthal plane (of Fig.~\ref{fig:v2-illustration}), roughly equal to $\vec{p}_{\bot, c} - \vec{p}_{\bot, d}$ and the azimuthal direction of the event plane ($\phi_{\q_T}$), given by $\q_T \simeq \vec{p}_{\bot, c} + \vec{p}_{\bot, d}$ (one can approximately view this as the angle between $p_c$ and the right pointing gray arrow in Fig.~\ref{fig:v2-illustration}). 

Using the events generated in our PYTHIA simulation, we compute the azimuthal angle distribution of hadrons with momentum  $p_T\!\geq\!5$~GeV,  relative to the direction $\phi_{\q_T}$.
We obtain the distribution shown in the left panel of Fig.~\ref{fig:template_event_plane_correlation}, where the green circles are for the typical intrinsic transverse momentum of $1$~GeV. The blue cross and red square points are results for $\langle k_\perp^2 \rangle^{1/2} = 3,5$~GeV respectively. With increasing $\langle k_\perp^2 \rangle^{1/2}$, one visually notes the appearance of an elliptic anisotropy.  
This notion can be quantified by fitting these points with the form 
\begin{equation}
	f(\phi) = 1 + 2 v_1 \cos(\phi-\phi_{\bm{q}_T}) + 2 v_2 \cos[2(\phi - \phi_{\bm{q}_T})]\;. \label{eq:template}
\end{equation}
The azimuthal anisotropy ($v_2$) increases, as the mean transverse momentum of the initial hard parton is increased. In the Supplement~\cite{supplemental}, we demonstrate using the same PYTHIA simulation, that a larger transverse energy ($E_T$), at forward rapidity, is produced with increasing $\langle k_\perp^2 \rangle$, leading to the identification of these as more central events. Thus, ``more central" events, have a larger initial intrinsic $\langle k_\perp^2 \rangle$ and a resulting higher $v_2$.

In addition to the $\cos(2\phi)$ correlation from polarization dependent cross sections ($d\Delta \sigma$)~\cite{Boer:2010zf,Pisano:2013cya,Conway:1989fs,NA10:1986fgk,EuropeanMuon:1983tsy, Barone:2009hw}, spin independent cross sections (i.e., $\sigma^0$) also produce angular asymmetries: Partonic cross sections will contain terms such as $\sfrac{\hat{s}}{\hat{t}}$ or $\sfrac{\hat{s}}{\hat{u}}$. Considering the $\sfrac{1}{\hat{t}}$ for the special case that the incoming parton has $p_a = \left[p_a, p_a \theta_a, 0 ,  p_a ( 1 - \theta_a^2/2 ) \right]$ (with $\theta_a \!\rightarrow\! 0$), and the outgoing parton has $p_c = \left[p_c, p_c \cos(\phi_c) , p_c \sin (\phi_c), 0  \right]$ (at mid-rapidity), we obtain, 
\begin{eqnarray}
	&\frac{\hat{s}}{\hat{t} } 
	= \frac{-\hat{s}}{ 2 p_c \cdot p_a } = \frac{-\hat{s}}{ 2p_c p_a [ 1 - \theta_a \cos(\phi_c) ] }\\
	&= \frac{-\hat{s}}{ 2p_c p_a } \left[1+\frac{\theta_a^2}{4} + \theta_a \cos(\phi_c) + \frac{\theta_a^2 \cos (2\phi_c)}{4} + \ldots \right]. \nonumber
\end{eqnarray}
Thus, the inclusion of initial state transverse momentum, leading to a non-zero $\theta_a \simeq \sfrac{k_{\perp,a}}{p_a}$, results in a variety of anisotropies $v_1 [\cos(\phi_c)$], $v_2 [\cos(2\phi_c)]$ etc. The inclusion of spin-polarization dependent states, with transverse polarized quarks and/or linearly polarized gluons, leads to additional sources of asymmetries, which will dominate over these power suppressed terms at high $p_T$~\cite{supplemental}. Other power suppressed terms leading to azimuthal asymmetries have been pointed out before in deep-inelastic scattering and Drell-Yan processes~\cite{Cahn:1978se, Berger:1979xz}.

\paragraph{Phenomenology and comparison with data:}
In this Letter, comparisons will be made with the ATLAS data on the elliptic anisotropy $(v_2)$ in $p$-$p$ and $p$-$Pb$ collisions, presented in Refs.~\cite{ATLAS:2016yzd,ATLAS:2019vcm} respectively. 
The experimental data is limited to $p_T \!\lesssim\! 50$~GeV. 
Our calculations will only include the partonic process of $g+g \rightarrow g+g$ (both unpolarized and linearly polarized). Subdominant processes, that prevail at higher $p_T$, can be found in Ref.~\cite{Soudi:2023TA}.

Since the linearly polarized portion of the gluon correlator in Eqns.~(\ref{eq:CorrelatorPDF}-\ref{eq:CorrelatorFF}) flips the helicity between the matrix element and its complex conjugate, the only allowed scattering involves two linearly polarized gluons.
While scatterings of two linearly polarized gluons into unpolarized gluons (double Boer-Mulders' functions) are allowed, we will focus on the dominant contribution for the second momentum anisotropy $(v_2)$, which is the coupling of one Boer-Mulders' function with one Collins' function ($BM\otimes C$). In this case, Eq.~\eqref{eq:sigma} simplifies as,
\begin{eqnarray}
	&&
	\Sigma^{\rm BM\otimes C}
	{\mkern-4mu}
	=
	{\mkern-14mu}
	\sum_{\lambda=\pm1}
	{\mkern-12mu}
	H^{\bot (1)}(z,k_{\bot C})
	{\mkern-4mu}
	\left[h^{\bot (1)}(x_a,k_{\bot a}^2)
	f(x_b,k_{\bot b}^2)
	\delta_{\alpha,1}
	\right.\nonumber\\
	&&
	\left. 
		+ f(x_a,k_{\bot a}^2)h^{\bot (1)}(x_b,k_{\bot b}^2)\delta_{\alpha,-1}
	\right]
	{\mkern-4mu}
	\hat{M}^{\lambda,\lambda}_{\lambda,\lambda} \left(\hat{M}_{-\alpha\lambda,\alpha\lambda}^{\lambda,\lambda}\right)^* 
	,\label{eq:sigmaBMC}
\end{eqnarray}
where we define $h^{\bot (1)} \equiv (k_{\bot}^2/2M_p^2)h^{\bot}$ and $H^{\bot (1)} \equiv (k_{\bot}^2/2M_{\pi}^2)H^{\bot}$.
The color and spin averaged matrix elements (times complex conjugate) can be expressed as, 
\begin{eqnarray}
	&\hat{M}_{\lambda,\lambda}^{\lambda,\lambda} \left(\hat{M}_{-\lambda,\lambda}^{-\lambda,\lambda}\right)^*
	\!\!\!\! =
	g_s^4 \frac{N^2}{N^2-1}\frac{t^2+tu+u^2}{t^2} e^{4\lambda i(\phi_{ab}-\phi_{bc})},
	&\\
	&\hat{M}_{\lambda,\lambda}^{\lambda,\lambda} \left(\hat{M}_{\lambda,-\lambda}^{-\lambda,\lambda}\right)^*
	\!\!\!\! =
	g_s^4 \frac{N^2}{N^2-1}
	\frac{t^2+tu+u^2}{u^2}
	e^{4\lambda i(\phi_{ab}-\phi_{ac})},
	&
\end{eqnarray}
where, using partonic momenta in spherical coordinates $\p_i=(p_i,\theta_i,\phi_i)$, the phases are given by
\begin{eqnarray}
    	&&\!\!\!\!\tan\phi_{ij} = \tan\frac{\phi_j-\phi_i}{2} \!\left(\sin\frac{\theta_j+\theta_i}{2}\right) \!\!\bigg/ \!\!\left(\sin\frac{\theta_j-\theta_i}{2}\right)
	{\mkern-2mu}.~~~~
\end{eqnarray}

Constrains on the gluon correlator lead to a bound on the spin-dependent distributions known as the Soffer bound: $|h^{\bot (1)}(x, k_{\bot}^2)| \leq f(x, k_{\bot}^2)$ and $|H^{\bot (1)}(x, k_{\bot}^2)| \leq D(x, k_{\bot}^2)$ \cite{Soffer:1994ww,Mulders:2000sh}. While the Soffer bound only constrains the spin-polarization dependent distributions to be limited by the regular distributions, the actual magnitudes of the Boer-Mulders' and Collins' functions are poorly constrained. 
In this effort, we take the spin-polarized distribution to be positive and proportional to the spin-polarization independent distribution, i.e., 
\begin{equation}
    h^{\bot (1)}\!(x,k_{\bot}^2\!) \!=\!b \!\cdot\!\! f(x, k_{\bot}^2\!) \; ;  
    H^{\bot (1)}\!(x,k_{\bot}^2\!) \!=\!B \!\cdot\!\! D(x, k_{\bot}^2\!), \label{eq:soffer}
\end{equation}
where, $b$, $B$ are fractions assumed independent of $x,k_\perp$. 
    
Recently, parametrizations of hard scale ($Q^2$) dependent joint spin-independent TMDPDFs [$f(x,k_\perp^2,Q^2)$, where the $x$ and $k_\perp$ dependence is not factorized] have appeared~\cite{BermudezMartinez:2018fsv,BermudezMartinez:2021lxz}. These can be directly used for $p$-$p$ collisions. However, there are no parametrizations of un-factorized TMDPDFs including nuclear effects, that can be used for $p$-$Pb$ collisions. To obtain such a distribution, calculations of parton branching in the presence of multiple scattering will have to be carried out. In the absence of such calculations, we employ a simple two step model: The TMDPDFs and TMDFFs will be assumed to be factorized with a Gaussian \emph{Ansatz}~\cite{Gamberg:2005ip,DAlesio:2004eso}:
\begin{eqnarray}
  &f(x,k_{\bot}, Q^2) = \frac{4\pi}{\langle k_\bot^2\rangle} e^{-\sfrac{k_\bot^2}{\langle k_\bot^2\rangle}} f(x,Q^2)\;,  
\end{eqnarray}
and the mean transverse momentum $\langle k_\bot^2\rangle$ in $p$-$A$ collisions, for the hard parton in the proton, is enhanced by $A^{1/3}$ (from multiple scattering) compared to $p$-$p$ as, 
\begin{eqnarray}
    &\langle k_\perp^2 \rangle_{pA} \simeq A^{1/3} \langle k_\perp^2 \rangle_{pp}\;.
\end{eqnarray}

This extra transverse momentum originates in multiple scattering of the hard parton prior to the hardest scattering which generates the di-jet, with $\Delta \langle k_\perp^2 \rangle \!=\! \langle k_\perp^2 \rangle_{pA}\! -\!\! \langle k_\perp^2 \rangle_{pp} \simeq \hat{q} L$, where $\hat{q}$ is the transverse momentum transport coefficient in confined matter~\cite{Baier:1996sk}, and $L \propto A^{1/3} \! -\! 1 $ is the average length traversed prior to hard scattering.
This $A^{1/3}$ enhancement, demonstrated for spin-independent multiple scattering~\cite{Guo:1997it, Fries:2000da, Fries:2002mu, Majumder:2007hx}, has not been established for polarization dependent  scattering. 
While transverse momentum is enhanced by multiple scattering, we expect vanishing initial state energy loss. This can be immediately understood from the simple schematic that energy loss $\Delta E \!\propto\! \hat{q} L^2$~\cite{supplemental}. For a Lorentz contracted nucleus, a finite $\Delta \langle k_\perp^2 \rangle$ is consistent with $\Delta E\rightarrow 0$, ensuring the $R_{pA}$ remains unmodified.

We employ the nCTEQ parametrization~\cite{Kovarik:2015cma} for integrated PDFs [$f(x,Q^2)$] and leading order KKP~\cite{Kniehl:2000fe} for FFs.
We present results with a constant value of $\langle k_\bot^2\rangle_{pp} = 1$GeV$^2$ and $\langle k_\bot^2\rangle_{pA} = A^{1/3}$GeV$^2$.
While the $\langle k_\perp^2\rangle$ is still not $Q^2$ dependent, we present results with a range of values for the bound $b\cdot B$ ($0.1\leq b \cdot B \leq 0.4$) to encompass a variety of uncertainties. 
For the pion TMD fragmentation, we employ a constant value of $\langle k_\bot^2 \rangle^{1/2} = 0.8$GeV.
We have explicitly checked that all these formulations produce almost no change in the angle integrated spectrum~\cite{Soudi:2023TA}.

\begin{figure}
	\begin{center}
		\includegraphics[width=0.43\textwidth]{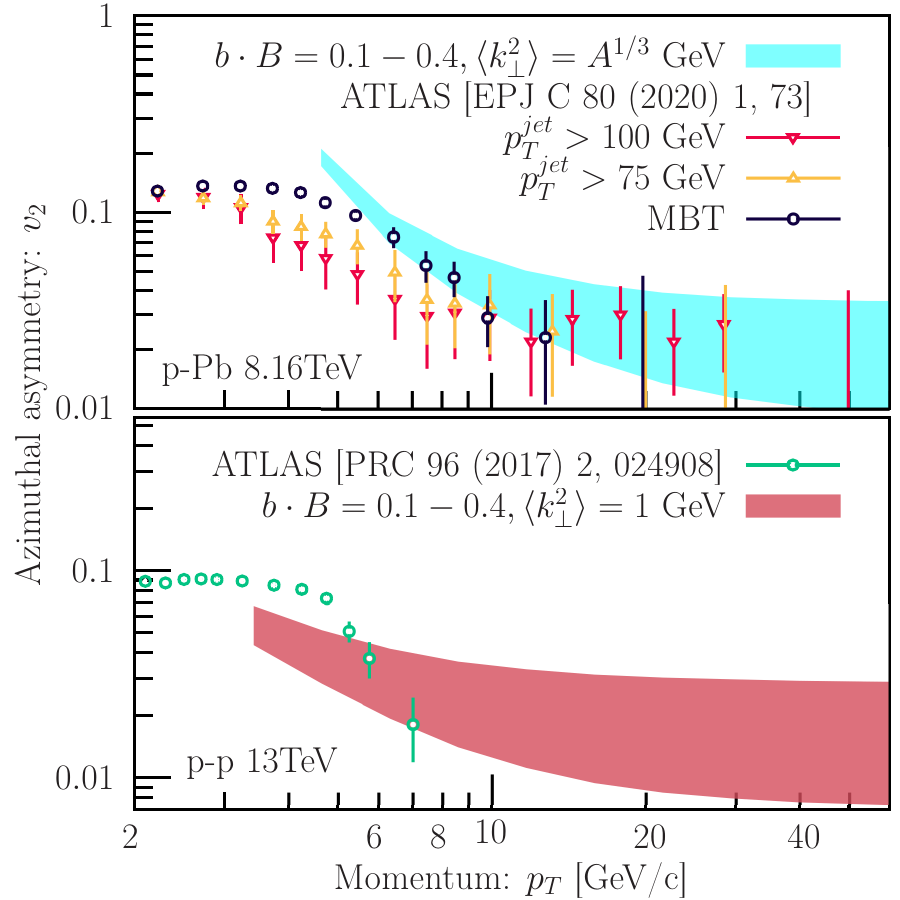}
        \vspace{-0.7cm}
	\end{center}
	\caption{Color Online: Azimuthal anisotropy coefficient $v_2$ as a function of pion transverse momentum $p_T$ for $p$-$p$ collisions at $\sqrt{s}=13$~TeV (bottom) and $p$-$Pb$ at $8.16$~TeV (top).
 }
	\label{fig:v2Band}
\end{figure}

The elliptic coefficient $v_2$ is obtained from the average:
\begin{equation}
\langle \cos[2(\phi_{\q_T} - \phi_{\pi})] \rangle \equiv \frac{\int d\phi_{\q_T} \cos [2(\phi_{\q_T} - \phi_{\pi})] \frac{d\sigma}{dP_T}}{\int d\phi_{\q_T}\frac{d\sigma}{dP_T}},   
\end{equation}
where, $\phi_\pi$ is the direction of the $\bm{\pi}$, and $\phi_{\q_T}$ is the direction of the reaction plane (Fig.~\ref{fig:v2-illustration},\ref{fig:template_event_plane_correlation}). 
Results for $p$-$p$ collisions at $\sqrt{s}=13$~TeV are shown in the bottom panel of Fig.~\ref{fig:v2Band}, where ATLAS results~\cite{ATLAS:2016yzd} are used to constrain the magnitude of the product of the TMDPDF and TMDFF to $10\%-40\%$ of the Soffer bound, i.e. 
$0.1 \leq b\cdot B \leq 0.4$~\eqref{eq:soffer}.
Using the same parameters, we obtain $v_2$ for $p$-$Pb$ collisions at $8.16$~TeV in the top panel of Fig.~\ref{fig:v2Band} and compare to ATLAS results \cite{ATLAS:2019vcm}.

We reiterate that the only difference between the calculations for $p$-$p$ and those for $p$-$A$ is the $A^{1/3}$ factor enhancing the $\langle k_\perp^2 \rangle$. The results in Fig.~\ref{fig:v2Band} include both polarization dependent and independent contributions. A decomposition into separate 
portions is presented in the Supplement~\cite{supplemental}, and shows that the polarization dependent piece dominates at $p_T \gtrsim 10$~GeV.

\paragraph{Summary:}
In this Letter, we have presented the first successful post-diction of an elliptic anisotropy in high $p_T$ particle production in $p$-$p$ and $p$-$Pb$ collisions with no modification to the angle integrated spectrum. We posit the source of this anisotropy to be entirely from the initial state, and \emph{not} from energy loss in the final state. 

Incoming PDFs and outgoing FFs can have intrinsic transverse momenta, which is enhanced for one PDF in $p$-$A$. Transverse momenta in the PDF selects out a preferred direction of the outgoing partons in the hard scatterings and correlates this with the net transverse momentum $\q_T$
of the two incoming hard partons. Conservation of momentum, within each incoming nucleon, correlates the net momentum of the bulk portion of the collision with the hard partonic scattering (Fig.~\ref{fig:v2-illustration}). Intrinsic transverse momentum allows for spin-polarization dependent partonic contributions, even with the incoming and outgoing hadrons being unpolarized (Boer-Mulders' and Collins' functions). These provide the dominant contributions to the anisotropy. These processes may also play a role in the high-$p_T$ anisotropy in $A$-$A$ collisions.

The calculations in this Letter can be improved upon:
Joint $x,k_\perp$ distributions, available for $p$-$p$, need to be calculated for $p$-$A$.
The use of factorized $k_\perp$ distributions requires further analysis and constraints from independent experimental data. The $A^{1/3}$ enhancement needs to be demonstrated for the case of spin-polarization dependent multiple scattering. Further, higher twist terms~\cite{Boer:2003cm} and contributions from Generalized TMDs~\cite{Boer:2018vdi} will have to be evaluated. 
While gluon-gluon scattering was dominant in the considered $p_T$ range, quark initiated processes will gain prominence at higher $p_T$, where mass effects may introduce additional corrections~\cite{Pisano:2013cya}. Finally, semi-analytical calculations need to be enhanced via extensive simulations, where correlations between the hard and soft sectors in these asymmetric collisions can be further investigated. 
All calculations in this Letter have assumed factorization between hard and soft sectors in high-$p_T$ hadron production. This is done with full knowledge that no generalized factorization theorem exists for processes with TMDs in $p$-$p$ collisions~\cite{Rogers:2010dm}. 

\paragraph{Acknowledgments:}
 Both authors are members of the JETSCAPE collaboration and have benefited from internal discussions. This work is supported in part by the U.S. DOE under Grant No.~{DE-SC0013460}.

\bibliography{ref}

\ifdefined\supplemental

\newpage
$\mbox{}$
\newpage

\section{Supplemental materials for: `Azimuthal Anisotropy at high transverse momentum in $p$-$p$ and $p$-$A$ collisions'}
\label{supplement}

\subsection{Results from an $x$-dependent $\langle k_\perp^2 \rangle$}

\begin{figure}[htb]
	\begin{center}
		\includegraphics[width=0.43\textwidth]{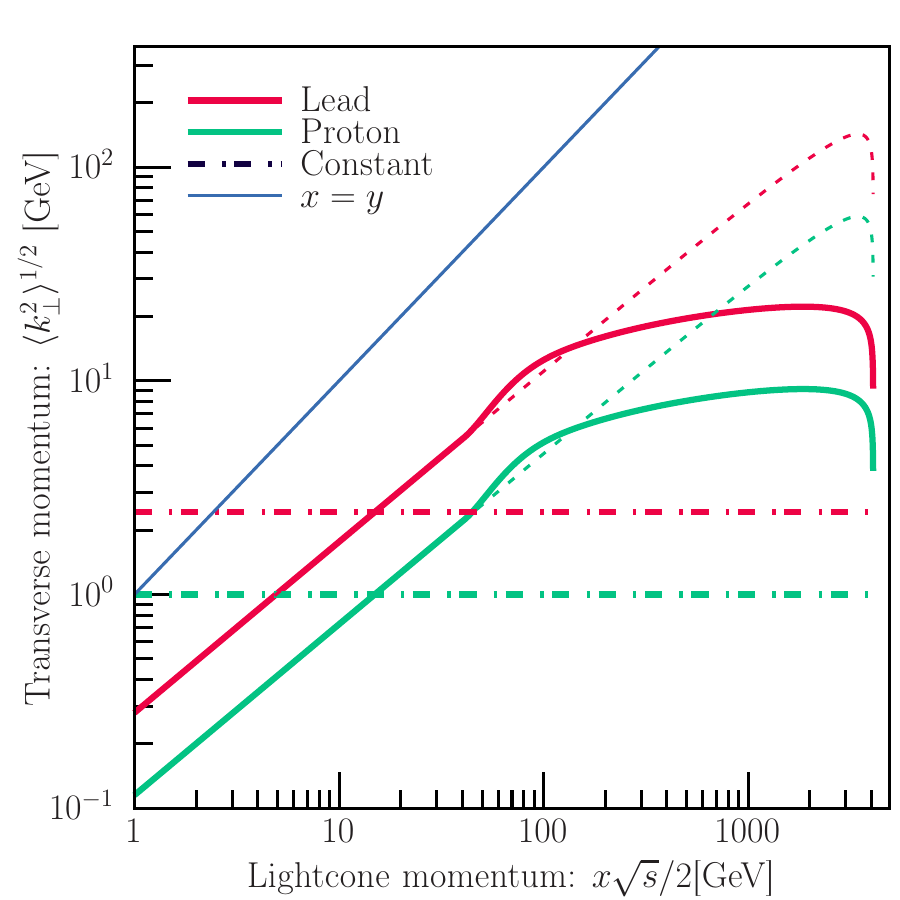}
	\end{center}
    \vspace{-0.5cm}
	\caption{Color Online: Average transverse momentum of gluons as a function of their lightcone momentum $x\sqrt{s}/2$.
	Green curves are results for proton-proton collisions, while red curves are multiplied by $A^{1/3}$ to represent proton-lead collisions.
	Dot-dashed lines display a constant $\langle k_\perp^2 \rangle$, while solid lines display the $x$-dependent $\langle k_\perp^2 \rangle$ from Eq.~(\ref{eq:kTFct}).
	}
	\label{fig:kTFct}
\end{figure}

 $x$-independent values of $\langle k_\bot^2 \rangle$ are not entirely physical: Small $x$-partons can end up with a larger transverse momentum than their light-cone momentum, as shown by the intersection of the dot-dashed lines with the blue solid line in Fig.~\ref{fig:kTFct}. An $x$-dependent $\langle k_\perp^2 \rangle$ \emph{ansatz} was presented in Ref.~\cite{DAlesio:2004eso}, which vanishes at both $x\!=\!0$ and $x\!=\!1$. However, it rises to large values in between, especially when enhanced by $A^{1/3}$ (shown by the light dashed lines in Fig.~\ref{fig:kTFct}). Calculations of the transport coefficient $\hat{q}$, in deconfined matter, using perturbative QCD, always contain a $\log{(E/E_0)}$ from the integration over phase space. $E$ is the energy of the hard parton and $E_0$ is a soft screening scale~\cite{Arnold:2008vd}. This motivates a modification of the $x$-dependent form of Ref.~\cite{DAlesio:2004eso} as~\footnote{For momentum fractions in the range $x_0 \leq x \leq 10 x_0$, we use a hyperbolic tangent $\left(\tanh\frac{x-x_0}{x_0}\right)$ to smoothly match the logarithmic behavior to the power-law.},
\begin{eqnarray}\label{eq:kTFct}
	&\langle k_\bot^2 \rangle^{1/2} (x) 
	= \langle k_\bot^2 \rangle_0^{1/2} \left((1-x)\sqrt{\frac{s}{s_0}}\right)^{0.15} A^{1/6} \nonumber\\
	 &\times
	\begin{cases}
	   \left(\frac{x}{x_0}\sqrt{\frac{s}{s_0}}\right)^{0.8}\;, &\text{ if } x \leq x_0\\
	   \left[1 + b_0 \log\left(\frac{x}{x_0}\sqrt{\frac{s}{s_0}}\right)\right]^{1/2} \;, &\text{ if } x > x_0.
	\end{cases}
\end{eqnarray}
We take $\langle k_\bot^2 \rangle_0\! =\! 5$ GeV$^2$, $x_0 \!=\! 10^{-2}$, $\sqrt{s_0}\!=\!8.16$~TeV and $b_0\!=\!2$ (solid red line in Fig.~\ref{fig:kTFct}). The constants are chosen to maintain the form of Ref.~\cite{DAlesio:2004eso} at $x \leq x_0$, turning over at larger $x$, and yielding a mean ($x$-integrated) $\langle k_\perp^2 \rangle$ comparable to the $x$-independent value~\cite{supplemental}. Strict $A^{1/3}$ scaling between $p$-$p$ and $p$-$A$ is maintained by setting $A=1$ for $p$-$p$ (solid green line in Fig.~\ref{fig:kTFct}).

\subsection{Contribution from polarized and unpolarized scatterings, and power corrections}

In the Letter above, we found that both unpolarized and polarized parton scattering can lead to azimuthal anisotropies at high-$p_T$. However, there is a difference: azimuthal anisotropies from unpolarized parton scattering are a power correction compared to polarized parton scattering. This is clearly demonstrated for the case of a constant $\langle k_\perp^2 \rangle^{1/2} =1$~GeV, for the case of $p$-$p$ collisions in Fig.~\ref{fig:v2pp-k-fixed-decomposition}, with $b\cdot B= 0.4$. 
One clearly notices the blue hatched area that represents the polarization independent term falling sharply compared to the polarization dependent term in solid green, with increasing $p_T$.

\begin{figure}[h]
	\begin{center}
		\includegraphics[width=0.43\textwidth]{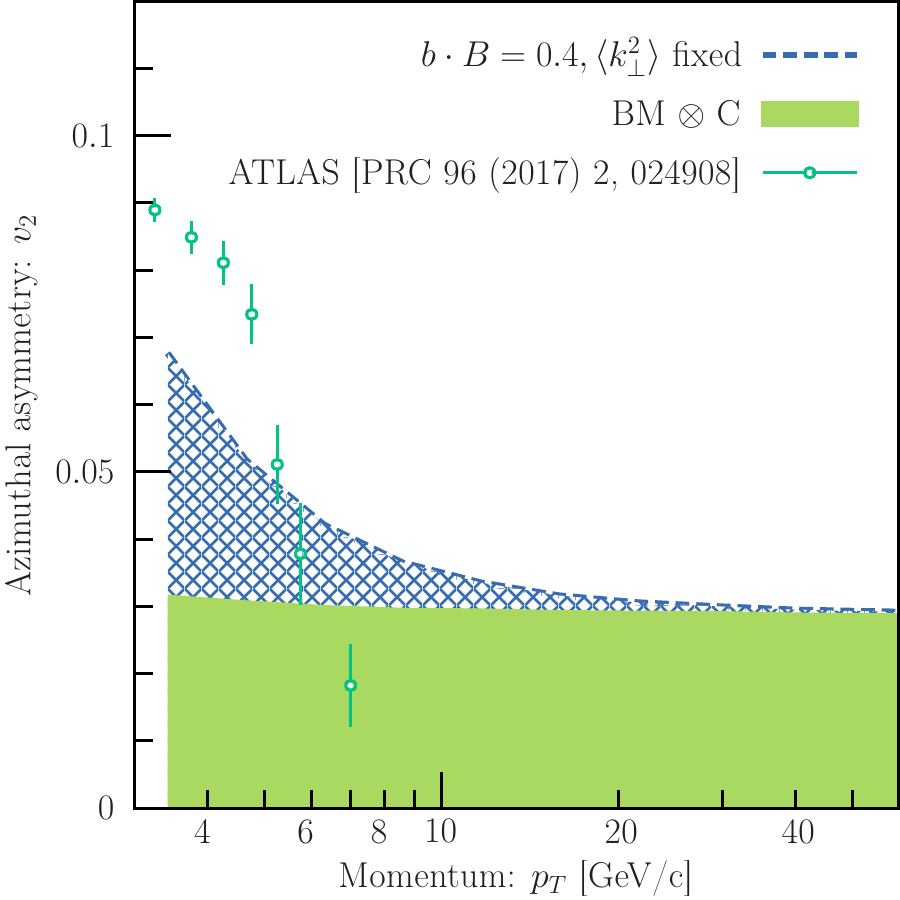}
        \vspace{-0.7cm}
	\end{center}
	\caption{Azimuthal anisotropy coefficient $v_2$ as a function of the pion transverse momentum $p_T$ for pp at $13$~TeV.
		The filled green represents the $\rm BM \otimes C$ contribution, the blue hatched represents polarization independent contributions.}
	\label{fig:v2pp-k-fixed-decomposition}
\end{figure}

Extending the fixed $\langle k_\perp^2 \rangle $ to $p$-$Pb$ by setting $\langle k_\perp^2 \rangle = A^{1/3}$~GeV, we obtain the plot in Fig.~\ref{fig:v2pPb-k-fixed-decomposition}, still with $b\cdot B=0.4$. The contribution from the polarization independent term is somewhat larger at $p_T \lesssim 10$~GeV. However, even in this case, it is suppressed at higher $p_T$ compared to the polarization dependent term.

\begin{figure}[h]
	\begin{center}
		\includegraphics[width=0.43\textwidth]{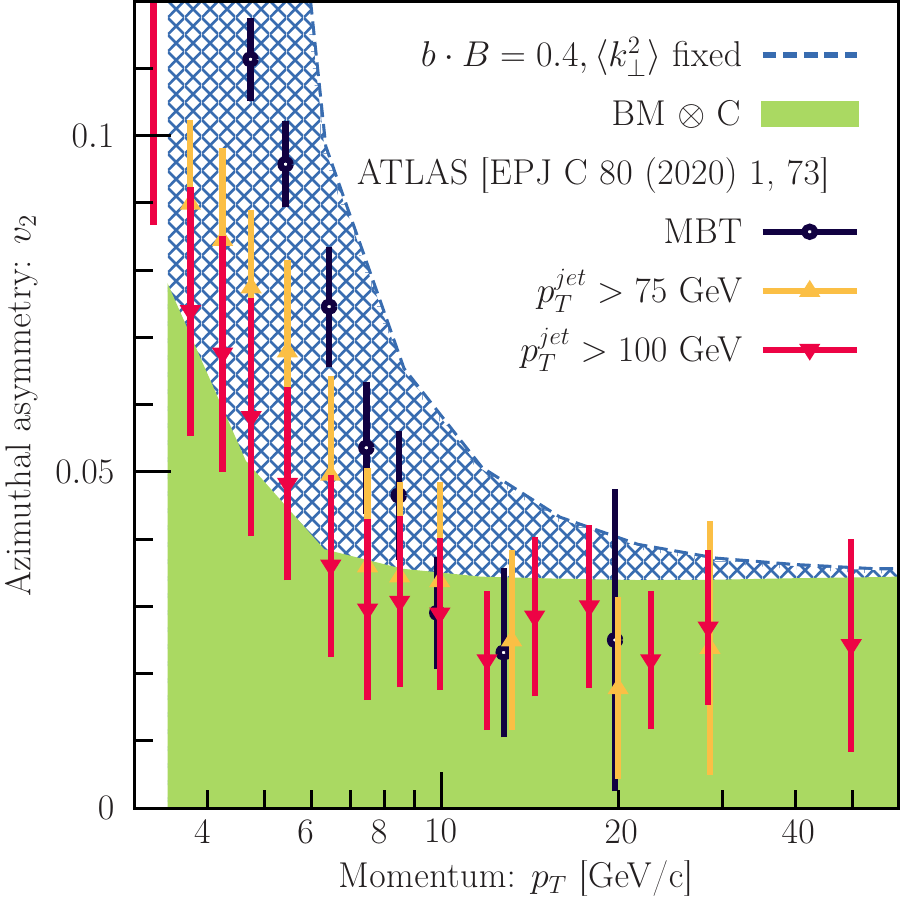}
        \vspace{-0.7cm}
	\end{center}
	\caption{Azimuthal anisotropy coefficient $v_2$ as a function of the pion transverse momentum $p_T$ for $p$-$Pb$ at $8.16$~TeV.
		The filled green represents the $\rm BM \otimes C$ contribution, the blue hatched represents polarization independent contributions.}
	\label{fig:v2pPb-k-fixed-decomposition}
\end{figure}

Using the $x$-dependent $\langle k_\bot^2\rangle$, reduces the suppression of the polarization independent terms, as they effectively produce a larger $\langle k_\perp^2 \rangle$ at larger $x$, and a much smaller $\langle k_\perp^2 \rangle$ at smaller $x$, 
as shown in Fig.~\ref{fig:kTFct}.  Using the form of $\langle k_\perp^2 \rangle^{1/2}$ from the previous subsection, we present the decomposition of the $x-$dependent $\langle k_\perp^2 \rangle$ contribution into the polarization dependent and independent terms for $p$-$p$ and $p$-$Pb$ collisions in Figs.~\ref{fig:v2pp-x-dependent-decomposition} and \ref{fig:v2pPb-x-dependent-decomposition} respectively.

\begin{figure}[h]
	\begin{center}
		\includegraphics[width=0.43\textwidth]{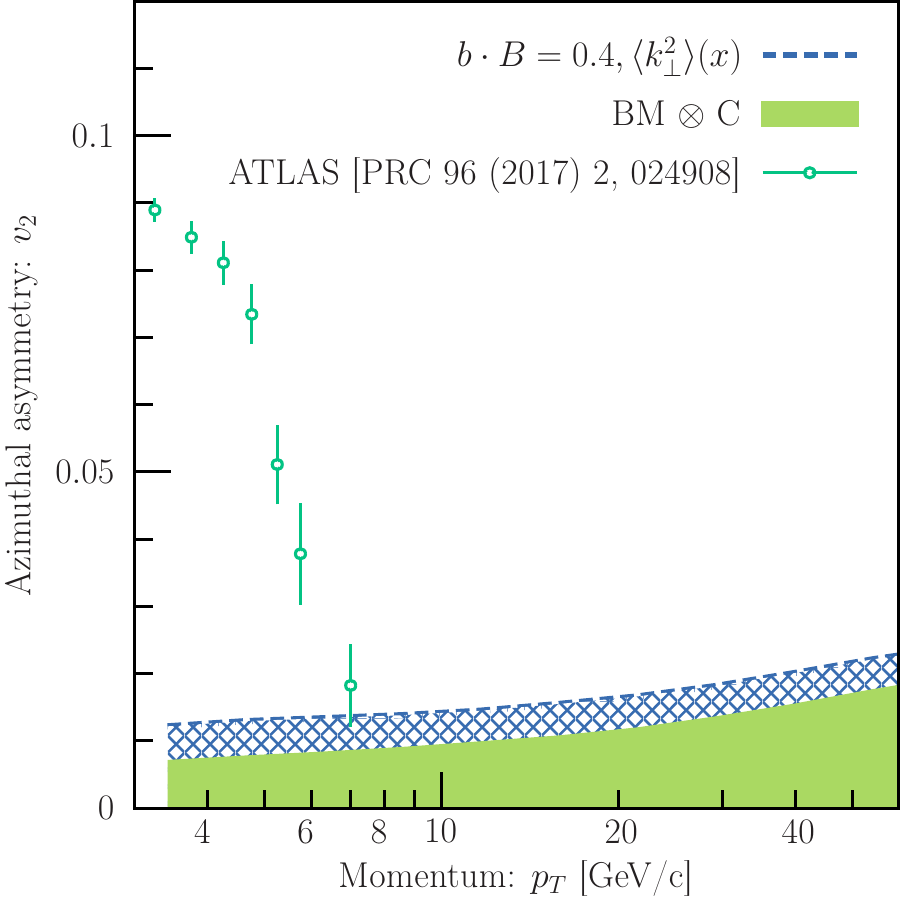}
        \vspace{-0.7cm}
	\end{center}
	\caption{Azimuthal anisotropy coefficient $v_2$ as a function of the pion transverse momentum $p_T$ for $p$-$p$ at $13$~TeV.
		The filled green represents the $\rm BM \otimes C$ contribution, the blue hatched represents polarization independent contributions.}
	\label{fig:v2pp-x-dependent-decomposition}
\end{figure}

In both Fig.~\ref{fig:v2pp-x-dependent-decomposition} and \ref{fig:v2pPb-x-dependent-decomposition}, the decomposition of contributions to $v_2$ from spin-polarization independent terms in blue hatched area and $\rm BM \otimes C$ terms in filled green for $p$-$p$ at 13TeV and $p$-$Pb$ at $8.16$~TeV show a similar trend.
Despite taking the magnitude of the product of TMD functions to be $40\%$ of the Soffer bound, we find the $\rm BM \otimes C$ contribution to be larger than the polarization independent term at $p_T \gtrsim 10$~GeV.

\begin{figure}[h]
	\begin{center}
		\includegraphics[width=0.43\textwidth]{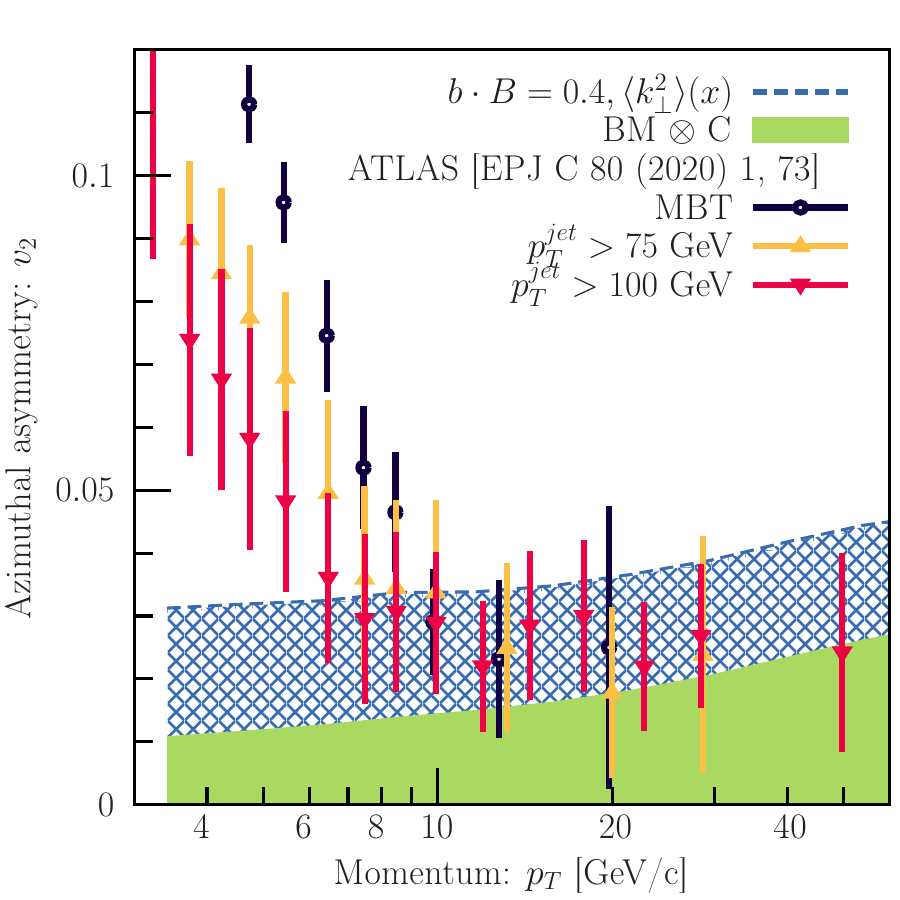}
        \vspace{-0.7cm}
	\end{center}
	\caption{Azimuthal anisotropy coefficient $v_2$ as a function of the pion transverse momentum $p_T$ for pPb at $8.16$~TeV.
		The filled green represents the $\rm BM \otimes C$ contribution, the blue hatched represents polarization independent contributions.}
	\label{fig:v2pPb-x-dependent-decomposition}
\end{figure}

\subsection{Centrality selection of events}
\begin{figure}[!h]
    \centering
    \includegraphics[width=0.43\textwidth]{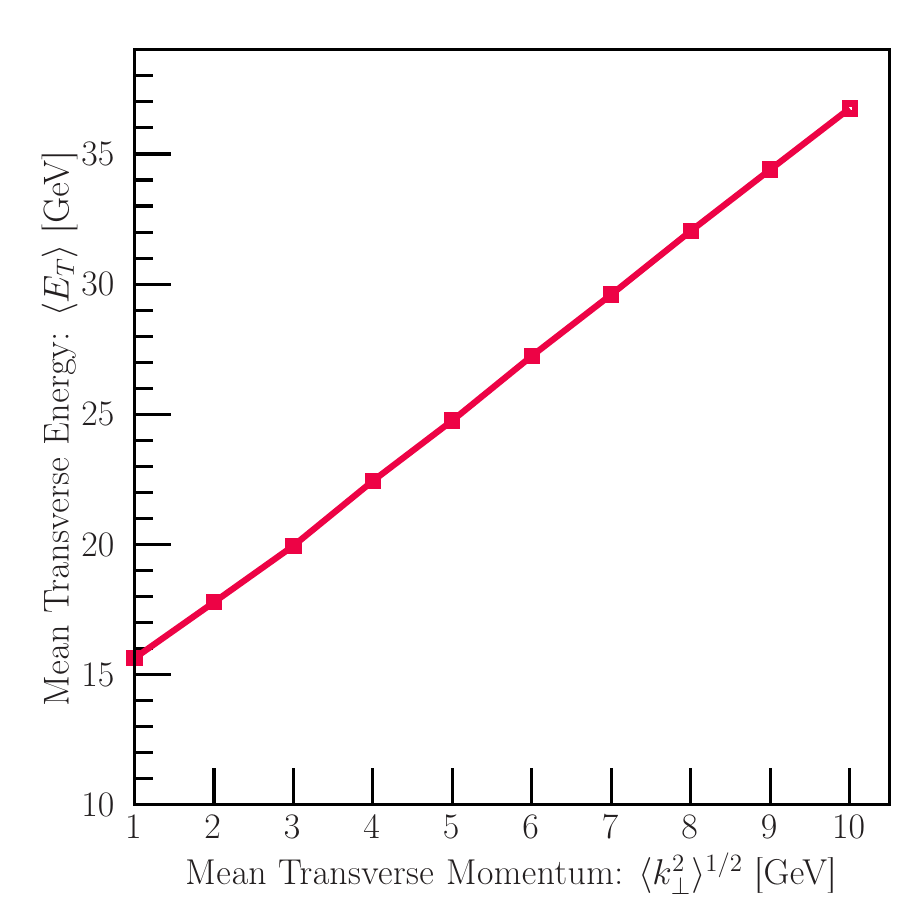}
    \caption{Average transverse energy $\langle E_T \rangle = E \sin\theta$ in the forward direction $3\leq|\eta|\leq 10$ as a function of the average transverse momentum of the initial hard parton $\langle k_\perp^2\rangle^{1/2}$}
    \label{fig:AverageET}
\end{figure}

Experimental studies of heavy-ion collisions perform centrality selections of the events based on the total transverse energy $E_T$ in the forward calorimeter (FoCal) (see e.g.~\cite{STAR:2004jwm,PHENIX:2003qra,ALICE:2010suc,CMS:2012zex,ATLAS:2012at}).
Events are considered to be more central when the transverse energy is large.
If we consider the total hard partonic scattering to be compensated by the remaining softer scatterings, a large transverse energy leads to a larger number of these partons deflected from the $z$-axis towards the FoCal rapidities.
We employ the Monte Carlo event generator PYTHIA 8 \cite{Bierlich:2022pfr} to study the effect of the initial transverse momentum on the correlation between hard and soft partons.

We set up a single hard scattering event without initial or final state parton shower and focus on the effect of initial intrinsic transverse momentum on the Lund model based fragmentation.
The Lund model relies on beam remnants, which are the partons that do not participate in the hard scattering but continue traveling along the $z-$direction.
For each remnant parton we sample a transverse momentum $\bm{k}_{\perp i}$ following a Gaussian distribution with an average momentum square $\langle k_\perp^2\rangle$ and random transverse direction.
The total transverse momentum of all remnants $\bm{q}_T=\sum_i \bm{k}_{\perp i}$ is compensated by applying a Lorentz boost to the center of mass frame of the hard scattering with the velocity 
\begin{align}
	\bm{v} = \frac{\bm{q}_T}{\sqrt{(E_a + E_b)^2 + \bm{q}_T^2}}\;.
\end{align}
By applying a boost to the center of mass frame, we ensure that the scattering cross-section is not altered.
However, this will entail that the initial hard parton will acquire a slightly different mean transverse momentum than the beam remnants.
We ignore this effect as it should not alter our conclusions.

We compute the transverse energy within a range of forward rapidities between $3 \leq |\eta | \leq 10$, 
\begin{align}
	E_T = \sum_i E_i \sin\theta_i\;,
\end{align}
where $E_i$ is the energy of the final state hadron and $\theta_i$ is the angle between the hadron momentum and the $z$-axis.
Only hadrons that lie within the selected pseudo-rapidity range are included in the sum above. 
The average transverse energy $\langle E_T \rangle$ is shown in Fig.~\ref{fig:AverageET} as a function of the average transverse momentum of the initial hard parton $\langle k_\perp^2\rangle^{1/2}$.
We find a clear linear relation between the initial transverse momentum of the hard parton and the transverse energy in the forward direction.
We conclude that centrality selection based on the transverse energy in the forward direction, where larger transverse energy is indicative of a more central event, leads to a selection of events where more central events are those with a large initial transverse momentum.


\subsection{Mean transverse momentum squared}
\begin{figure}
	\includegraphics[width=0.43\textwidth,angle=90]{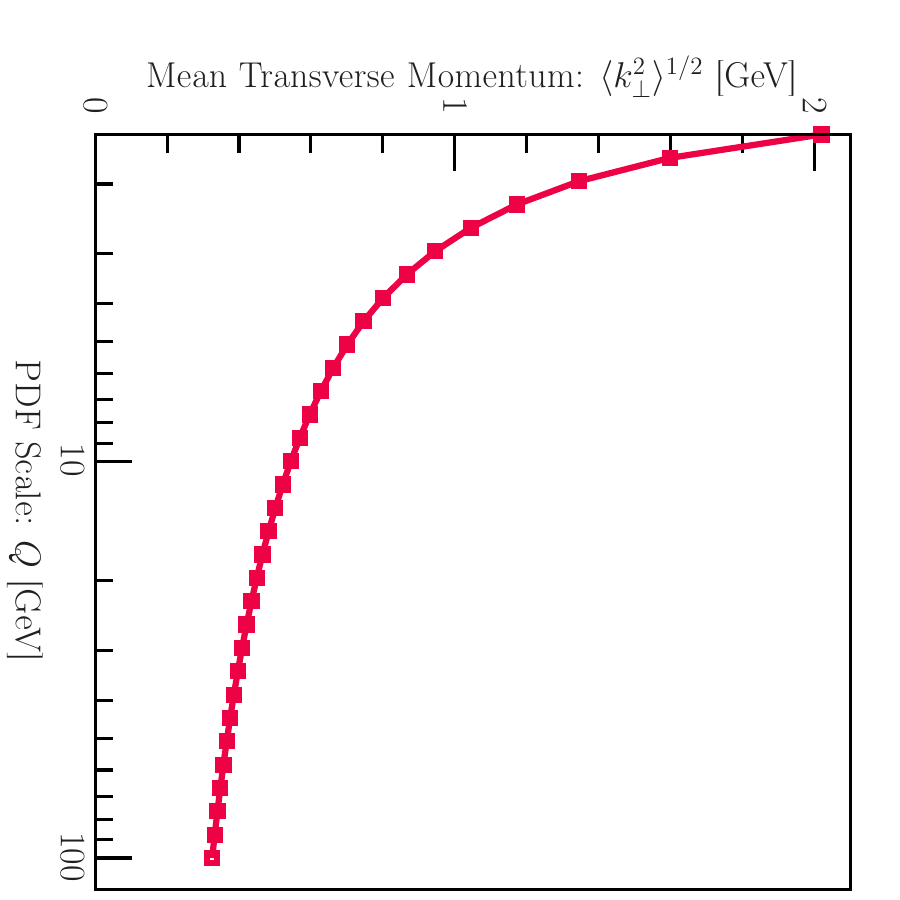}
	\caption{Mean transverse momentum squared $\langle k_\perp^2\rangle$ of the initial hard parton as a function of the scale Q of the PDF.}
	\label{fig:AveragekT}
\end{figure}
Due to the $x$-dependence of the mean transverse momentum squared in the Gaussian \emph{Ansatz} of Eq.~\eqref{eq:kTFct}, the factorization between transverse and longitudinal momenta distributions is broken.
Effectively, the mean transverse momentum of the partons is not directly the function $\langle k_\bot^2\rangle(x)$.
Instead, to obtain the effective $\langle k_\perp^2\rangle$ on needs to integrate the $x$-dependent $\langle k_\bot^2\rangle$ weighted by the PDF, as follows
\begin{align}
	\langle k_\perp^2\rangle(Q) = \frac{\int dx_a \langle k_\bot^2\rangle f(x_a,Q)}{\int dx_a f(x_a,Q)}\;,
\end{align}
where $Q$ is the scale where the PDF is evaluated.

Using the nCTEQ PDF for gluon in the proton, we obtain the mean transverse momentum squared of the initial hard parton as a function of the scale $Q$ in Fig.~\ref{fig:AveragekT}.
We see that, effectively, the transverse momentum does not exceed $2$GeV at low scales and as the scale increases the transverse momentum plateaus at around $\sim 0.6$GeV.

\subsection{Energy Loss from initial state scattering.}

The probability of radiating a gluon induced by initial state scattering prior to a hard event was calculated in Ref.~\cite{Fries:2000da}, the mirror calculation for a gluon produced by final state rescattering was carried out in Ref.~\cite{Wang:2001ifa}. Using a formula derived from their expressions, and including the interference of double hard and hard-soft terms, we obtain the probability for a quark with energy-momentum $(Q/2,0,0,Q/2)$ to radiate a gluon, prior to hard scattering, as 
\begin{equation}
\frac{dN}{dy}\!\! = \!\!\int\limits_0^{Q^2}\!\! \frac{\alpha_S}{2\pi}\frac{dl_\perp^2}{l_\perp^4} P(y) \!\!\!\int\limits_{-L^-}^0 \!\!\!d\tau^- \hat{q} \left[ 2 - 2 \cos \left( \frac{l_\perp^2}{Q} \tau^- \right) \right].  
\end{equation}
In the above equation, $\hat{q}$ is the transport coefficient in cold nuclear matter, $\tau^-$ is the light-cone location of the soft scattering, which ranges from the location of the hard scattering at $0$ to a light-cone distance $-L^-$. The splitting function $P(y)$, where $y$ is the light-cone momentum fraction carried by the radiated gluon can be approximated in the soft gluon limit as $P(y) \simeq 2/y$. The transverse momentum of the radiated gluon is $l_\perp$.

The energy lost by a single gluon emission from a quark with energy $Q/2$ can be estimated as
\begin{equation}
    \Delta E = \int_0^1 dy \left(y\frac{Q}{2}\right) \frac{dN}{dy}
\end{equation}
The integral of this expression yields a $\Delta E \sim {L^-}^3$ at very small distances $L^- \ll 1/Q$ and a $\Delta E \sim {L^-}^2$ at larger distances. As a result, for a Lorentz contracted nucleus in the lab frame, a parton traverses $L^-_{Lab} \!\!= \!\!L^-/\gamma_{Lab}$ with $\hat{q}_{Lab}\! =\! \gamma_{Lab} \hat{q}$, where $\gamma_{Lab}$ is the boost factor. Thus, for finite $\Delta \langle k_\perp^2 \rangle = \hat{q} L^- = \hat{q}_{Lab} L^-_{Lab}$ and $\gamma_{Lab} \gtrsim 1000$, one has a $\Delta E\rightarrow 0$, ensuring the angle integrated cross section (and $R_{pA}$) remains unmodified. 
\fi

\end{document}